\documentstyle[12pt,a4]{article}

\def\getup{\vspace{-1.1\baselineskip}}

\def\degres{\mbox{$^\circ$}}      

\def\lsim{\mbox{$\stackrel{_<}{_\sim}$}} 

\def\bec{\begin{center}}
\def\eec{\end{center}}

\def\beq{\begin{equation}}
\def\eeq{\end{equation}}

\renewcommand{\frac}[2]{{{\displaystyle #1}\over{\displaystyle #2}}}

\def\eV{\mbox{ eV}}      

\def\MeV{\mbox{ MeV}}

\def\BHt{\mbox{$^{8}$B}}

\def\calA{\mbox{$\cal A$}}     

\def\calR{\mbox{$\cal R$}}
\def\calS{\mbox{$\cal S$}}

\def\hath{\mbox{${\hat{h}}$}}

\def\dms{\mbox{$\Delta m^2$}}     

\def\SdTvS{\mbox{$\sin^2 2\theta_V$}}
\def\STvS{\mbox{$\sin^2 \theta_V$}}

\def\ThetaV{\mbox{$\theta_v$}}

\def\numt{\mbox{$\nu_{\mu(\tau)}$}} 
\def\nue{\mbox{$\nu_e$}}            
\def\num{\mbox{$\nu_\mu$}}          
\def\nut{\mbox{$\nu_\tau$}}         
\def\nutwo{\mbox{$\nu_2$}}          
\def\TYear{\mbox{$T_{y}$}}

\def\rhoR{\mbox{$\rho_R$}}      

\def\Ye{\mbox{$Y_e$}}           

\def\EDms{\mbox{$E_\nu/\Delta m^2$}} 
\def\Enu{\mbox{$E_\nu$}}         
\def\Te{\mbox{$T_e$}}            
\def\TeTh{\mbox{$T_{e,th}$}}     
\def\SeZr{\mbox{$\calS_{0}$}}       
\def\Ses{\mbox{$\calS^s$}}          
\def\SeD{\mbox{$\calS^D$}}          
\def\Res{\mbox{$\calR^s$}}          
\def\ReZr{\mbox{$\calR_{0}$}}       
\def\ReD{\mbox{$\calR^D$}}          
\def\AsymPs{\mbox{$\calA_P^s$}}     
\def\AsymSs{\mbox{$\calA_{D-N}^s$}} 
\def\AsymSC{\mbox{$\calA_{D-N}^C$}} 
\def\AsymRs{\mbox{$A_{D-N}^s$}}     
\def\AsymRN{\mbox{$A_{D-N}^N$}}     
\def\AsymRC{\mbox{$A_{D-N}^C$}}     
\def\AsymRM{\mbox{$A_{D-N}^M$}}     
\def\deltaSs{\mbox{$\delta \calS^s$}}   
\def\Ps{\mbox{${\bar{P}}_\odot$}}       
\def\PTot{\mbox{$P_{\oplus}$}}          
\def\PeTw{\mbox{$P_{e2}$}}              
\def\APeTw{\mbox{$<\PeTw>$}}            
\def\APeTws{\mbox{$<\PeTw>^s$}}         
\def\APeTwN{\mbox{$<\PeTw>^{N}$}}       
\def\APeTwC{\mbox{$<\PeTw>^{C}$}}       
\def\APeTwM{\mbox{$<\PeTw>^{M}$}}       
\def\TResids{\mbox{$T_{res}^s$}}      
\def\TResidD{\mbox{$T_{res}^D$}}      
\def\TResidN{\mbox{$T_{res}^N$}}      
\def\TResidC{\mbox{$T_{res}^C$}}      
\def\TResidDC{\mbox{$T_{res}^{DC}$}}  
\def\TResidM{\mbox{$T_{res}^M$}}      
\def\DAY{\mbox{\em{Day}}}                        
\def\night{\mbox{\em{Night}}}                    
\def\core{\mbox{\em{Core}}}                      
\def\mantle{\mbox{\em{Mantle}}}                  
\def\deepcore{\mbox{\em{Deep - Core}}}           
\def\hatdelta{\mbox{$\hat{\delta}$}}             
\def\RSE{\mbox{$R$}}                   
\def\daynight{D-N}                        
\def\SK{Super - Kamiokande}
\def\maxim{\mbox{max}}                    

\def\deg{\degres}  

\def\CdTv{\mbox{$\cos 2 \theta_V$}}

\def\NA{Nadir angle}

\def\ltap{\lsim}  

\def\electron{\mbox{$e^-$}}

\def\FigPeTw{1}
\def\FigAPeTw{2}
\def\FigSDist{3}
\def\FigContour{4}
\def\FigContourFive{5}

\begin{document}
\sloppy


\vspace{4cm}
\bec{\huge Comprehensive Study of the Possibility of\\
       Day - Night Effect Enhancement}\eec

\vspace{2 \baselineskip }

\large

\bec
Michele Maris\\
Pavia University - Italy\\
and\\
SISSA Trieste - Italy
\eec

\vspace{1cm}

\bec {\bf{Talk held at the:}}\\
 {\em Fourth Solar Neutrino Conference}\\
April 8 - 11, 1997\\
Heidelberg, Germany \\
\eec 

\vspace{1cm}

\bec
\abstract{
\large
\noindent
A set of quantitative predictions for the \daynight\ asymmetry in the \SK\ 
detector is presented. 
For these predictions, neutrino events are collected in ``samples'' 
defined by the 
trajectories in the Earth of the corresponding solar neutrinos.
The more important samples considered here are:
the full sample of neutrinos detected at night, 
the sample of night neutrinos which cross the
Earth core and the sample of night neutrinos which does not cross the 
Earth core.
For energy integrated event rates, the \daynight\ asymmetry for core crossing
neutrinos is up to six times bigger than the \daynight\ asymmetry 
for the full sample of night neutrinos.
When the reduction in statistics is considered, an effective enhancement up to 
a factor $\approx 2.3$\ is obtained. 
The selection of core crossing neutrinos may be relevant for the 
recoil - \electron\ spectra too.
From these results it is reasonable to expect the \SK\ detector be able 
to constrain the small 
mixing angle solution of the solar neutrino problem 
($\sin^2 2 \theta_v \leq 0.01$) through the \daynight\ effect.
Furthermore, it is proposed that the \daynight\ effect may limits the total
flux of \BHt\ neutrino produced in the Sun.
At last, the sensitivity of these predictions to the Earth model uncertainties 
is quantitatively discussed.
 }
\eec

\newpage

\section{Introduction}
This talk reports the main results from a research conduced by the speaker 
in collaboration with
Serguey T. Petcov  
(SISSA - Trieste, Italy, also at Institute of Nuclear
Research and Nuclear Energy, Bulgarian Academy of Sciences,
1784 Sofia, Bulgaria)
and Q.Y. Liu 
(SISSA - Trieste, Italy)
about the \SK\ sensitivity
to the ``day-night effect'' \cite{ArticleI, ArticleII}.
The research was aimed by two main purposes: to produce quantitative 
predictions for the day-night effect as seen by the \SK\ detector,
to study the enhancement in the day-night asymmetry  
obtained  selecting those neutrino events due to solar 
neutrinos which cross the Earth core in their way to the detector.

The observation of the Earth or day-night (D-N) effect would be a proof 
of the validity of the MSW solution of the solar neutrino problem,
since the Earth effect in neutrino propagation is a direct consequence of 
neutrino oscillations in matter \cite{MSW:Original}. 
In its simplest version the MSW mechanism involves matter - enhanced 
two-neutrino transitions of the solar \nue\ into an active neutrino 
\num\ or \nut, $\nue \rightarrow \nu_{\mu(\tau)}$, while the \nue\ 
propagates in the Sun.
In this case 
the $\nue \rightarrow \nu_{\mu(\tau)}$\ transition probability depends 
on two parameters: \dms\ and \SdTvS, where $\dms > 0$\ is the 
neutrino mass squared difference and $\ThetaV$\ is the angle 
characterizing the neutrino mixing in vacuum.
If $\dms \ne 0$\ and $\ThetaV \ne 0$, the neutrino propagation is sensitive to 
the matter
distribution along the propagation path and the probability to detect at
Earth a \nue\ produced in the Sun is a function of the detection time $t$
because at night the Sun is below the horizon and the solar neutrinos cross
the Earth reaching the detector, so that further $\nue \rightarrow \numt$\ 
transitions occur.
This can lead to a difference or ``asymmetry'' between the signals caused by 
the solar 
neutrinos in a solar neutrino detector during the day and during the night, 
i.e., to a D-N asymmetry in the signal.
No other mechanism of depletion of the solar \nue\ flux proposed so far 
can produce such an effect.

For the Bahcall and Pinsonneault (1995) solar model 
and assuming MSW $\nue \rightarrow \numt$\ transitions
the solar neutrino data can be described for values of the parameters
\dms\ and \SdTvS\ belonging to two intervals \cite{Krastev:Petcov:1996}:
$  3.6 \times 10^{-6} \, \mbox{ eV}^2 \, \ltap \, \Delta m^2 \, \ltap \, 
9.8\times 10^{-6}\mbox{ eV}^2$,
$  4.5 \times 10^{-3} \, \ltap \, \SdTvS \, \ltap \, 1.3\times 10^{-2}$;
or 
$ 5.7 \times 10^{-6} \mbox{ eV}^2 \, \ltap \, \Delta m^2 \,
                                     \ltap\, 9.5\times 10^{-5} \mbox{ eV}^2$,
$    0.51 \,\ltap \, \SdTvS \, \ltap\, 0.92$.
These intervals 
correspond to the nonadiabatic (small mixing angle) and to the adiabatic
(large mixing angle) solutions of the solar neutrino problem.
When theoretical uncertainties in solar neutrino flux components
are considered, larger ``conservative'' regions 
\cite{MSW2,Krastev:Petcov:1996}:
$  3.0 \times 10^{-6} \mbox{ eV}^2 \,\ltap\, \Delta m^2 \,\ltap\, 1.2\times 
10^{-5}\mbox{ eV}^2$,
$  6.6 \times 10^{-4} \,\ltap\, \SdTvS \,\ltap\, 1.5\times 10^{-2}$;
or
$    7.0 \times 10^{-6} \mbox{ eV}^2 \,\ltap\, \Delta m^2 \,\ltap\, 1.6\times 
10^{-4} \mbox{ eV}^2$, 
$    0.30 \,\ltap\, \SdTvS \,\ltap\, 0.94$.

To reduce the complexity of the required calculations
experimental details such as the background subtraction and the systematic
errors were neglected, also continuous data taking, with no 
interruptions, were assumed.
The only experimental constraints which must be 
considered are:
the 5 MeV threshold in the recoil-\electron\  kinetic energy \TeTh,
and the total yearly event rate for the given threshold energy, 
which is expected to be $\approx 10^4$\ 
events/year which allows a statistical error $\approx 1\%$\ for one year of data.
In this way, predictions presented here are for an idealized 
\SK\ detector at its best performances. They should be considered the 
starting point to produce predictions which take in account background,
systematic errors and the data taking history.
Apart from the Bahcall and Pinsonneault (1995) solar model 
\cite{BP95}, and the Stacey (1977) Earth model, other source of data are
\cite{Bahcall:etal:1996}.

\section{Calculating the Earth Effect}
 The probability for the process $\nue \rightarrow \nue$\ 
for a solar neutrino propagating in the Sun and Earth is 
given by \cite{MS:DNEFF:1986, Baltz:Weneser:1994}\
\getup\bec\beq\label{eq:neutrino:survival:proba}
    \PTot(\nue \rightarrow \nue) = \Ps(\nue \rightarrow \nue) + 
             \frac{1 - 2\Ps(\nue \rightarrow \nue)}{\CdTv} 
      \, (\PeTw - \STvS),
\eeq\eec

\noindent
where $\Ps(\nue \rightarrow \nue)$\ is the average probability of solar \nue\ 
survival in the Sun and \PeTw\ is the probability
of the $\nutwo \rightarrow \nue$\ transition after the \nue\ have left the Sun.
The probability \PeTw\ is not
a constant in time because it is a function of the trajectory followed by the
neutrinos crossing the Earth, determined by the instantaneous
apparent position of the Sun in the sky.
Given the fact that data are taken over a certain interval of time,
the instantaneous \PeTw\ in eq. (\ref{eq:neutrino:survival:proba}) has to be
replaced with its time averaged, \APeTw, which requires the computation of
the solar position in the sky.

For the scopes of \daynight\ effect calculations it is possible to assume for 
the Earth structure a spherical symmetry, so that the electronic density is 
radially distributed.
In this case the neutrino trajectory and the serie of geological stratifications
it crosses reaching the detector 
are defined by the {\em Nadir Angle}\ for the
Sun \hath, which is the angle between the direction of the
Earth center (Nadir) 
and the direction of the Sun as seen by the detector.
From the detection time $t$ of a neutrino event it
is possible to recover $\hath(t)$\ allowing the selection of events in 
accord with the trajectory of the incoming neutrinos 
to form a set of five ``samples'' labeled:
\DAY, \night, \mantle, \core\ and \deepcore. 
All the quantities which refers
to a specified sample (probabilities, spectra, etc.) 
are labeled by an index $s=D$, $N$, $C$, $M$, $DC$\ 
for \DAY, \night, \core, \mantle\ and \deepcore\ samples respectively. 
Samples are defined as follow:
the \DAY\ sample is formed by all neutrino events detected at day; 
the \core\  sample by those neutrinos which cross the Earth core 
($\hath \leq 33.17\deg$); 
in the \mantle\ sample neutrinos does not cross the Earth core;
($\hath \geq 33.17\deg$);
the \night\ sample is defined merging the  \core\ and  \mantle\ sample;
at last in the \deepcore\ sample neutrinos cross the inner $2/3$\ of core.

The computation of $\Ps(\nue \rightarrow \nue)$\
is accomplished following the prescriptions 
derived in ref. \cite{Petcov:1988}.
For a given Earth model and 
a fixed set of values for the parameters \hath, \SdTvS\ and \EDms,
\PeTw\ is computed solving numerically 
the ordinary differential equations for the neutrino propagation in the Earth.
The differential equation is integrated through a 4$^{th}$\ order Runge - Kutta
algorithm with an adaptive step size.
A set of tests assures this procedure
is stable and accurate at the $\approx 10^{-5}$\ level.
Furthermore,
all over the program checks were made to assure consistency and proper 
numerical error control, based on the comparison
of results obtained by the {\tt FORTRAN} code with results obtained 
by a {\tt MATHEMATICA} code. 
The comparison shows relative differences $\approx 10^{-5}$\ in all the 
relevant cases. In conclusion the numerical error is under control and 
quite negligible.

Apart from the \DAY\ sample, 
each sample $s=D$, $N$, $C$, $DC$\ requires a 
different calculation of the time average probability \APeTws\ from the
instantaneous probability \PeTw.
For a sample $s=D$, $N$, $C$, $DC$\ 
defined by the nadir angle interval $[\hath_1^s,\hath_2^s]$, and 
obtained over an exposure time interval $[T_1^s, T_2^s]$
(assumed here to last from January $1^{st}$ 1996 to January $1^{st}$ 1997)
the averaged \PeTw\ probability is defined as:
\getup\bec\beq\label{eq:APeTw:Time}
\APeTws =
  \frac{1}{\TResids}
{\displaystyle
  \int_{T_1^s}^{T_2^s} dt\, 
   \frac{\hatdelta(\hath_1^s \leq \hath(t) \leq \hath_2^s)}{\RSE^2(t)} \,
    \PeTw(\hath(t)) },
\eeq\eec

\noindent
where:
$\hatdelta$\ is 0 for \hath\ outside the nadir angle interval and 1 for 
\hath\ inside it; 
$\RSE(t)$\ is the Sun - Earth distance expressed in units of
the mean distance (the Astronomical Unit $R_0 = 1.4966 \times 10^8$\ km);
and 
\getup\bec\beq\label{eq:RTime}
  \TResids = {\displaystyle \int_{T_1^s}^{T_2^s} dt \, 
              \frac{\hatdelta(\hath_1^s \leq \hath(t) \leq \hath_2^s)}
                   {\RSE^2(t)} }
\eeq\eec

\noindent
is the {\em residence time}, i.e., the time spent by the Sun in 
$[\hath_1^s,\hath_2^s]$. 
In the absence of neutrino oscillations the total fraction of solar neutrinos 
induced events in $[\hath_1^s,\hath_2^s]$ is 
$\TResids/\TYear$, where $\TYear =  (T_2^s -  T_1^s)$ 
in this case is the year length.
For the \SK\ detector the times over one year of data taking are: 
$\TResidN/\TYear  = \TResidD/\TYear =  0.5$, 
$\TResidC/\TYear  = 0.071$, 
$\TResidM/\TYear  = 0.429$, 
$\TResidDC/\TYear  = 0.0394$.
With a statistics of $10^4$ events this corresponds to a relative statistical
error of about 1.4\% (\night\ or \DAY), 3.8\% (\core), 1.5\% (\mantle), 5.0\% 
(\deepcore).
The residence time is also relevant because it weights the
contributions to $\APeTwN$\ from each geophysical structure  and from the 
 error source.

Many types of systematic errors can affect the predictions for the \daynight\
asymmetry. Since the initial hypothesis,
the most relevant error sources for these calculations 
are related to the Earth model and 
the apparent solar motion \cite{ArticleI}.

Earth model uncertainties introduce a relevant error source 
(especially for the core)
since they 
affects the Earth electronic density $n_e \propto \Ye \rho$,
where \Ye\ is the fraction of electrons per nucleon and $\rho$\ is the matter
density (in Dr/cm$^3$), 
the combined $n_e$\ uncertainty is $\approx 7\% \,\div\, 10\%$
\cite{ArticleII}. 
Its effect was studied repeating the calculations 
taking the $\rho$\ distribution \cite{Stacey:1977}\ but for two different
hypothesis about \Ye\ in the core: 
$\Ye=0.467$\ which is representative of current knowledge and 
$\Ye=0.5$\ 
which combines all the uncertainties in Earth models.
In both the cases it is assumed $\Ye=0.5$\ in the mantle
\cite{ArticleI, ArticleII}.

The apparent solar motion is quite complex and 
approximations are required to save computing time.
The limit in the accuracy for the Sun's tracking is fixed by 
the angular diameter 
of the \BHt\ neutrino production region $\approx 3$ arcmin
\cite{ArticleI}.
There are two kinds of approximation 
made which are relevant for the Earth effect. 
In the first the Earth motion around the Sun is 
circular (COA), in the second it is elliptical (EOA),
both of them 
are not exact models, but EOA  is very accurate for Earth effect prediction 
purposes.

The systematic errors in the Sun position produced by the COA and the EOA
affect (\ref{eq:APeTw:Time}) and (\ref{eq:RTime}) in two ways:
i) they change 
the time in which the Sun cross the 
boundaries of a given \NA\ interval, ii) they change \hath(t).
The first way is the less relevant, since it mainly affects the residence 
time,
which is the difference between two next crossing times, so that errors 
compensate.
The second way is more relevant, because 
it affects the length of the neutrino path in a given
geological structure.
The analysis shows that EOA introduces errors in the Sun position 
smaller than $0.01$\deg, while COA may introduce errors as large as few 
$0.01$\deg. 
Most of these errors are periodic in 
nature, and they should  average out.
However, since \PeTw\ is a function of the solar position and 
the use of different sampling schemes (as \core, \mantle, \night, etc.) are 
equivalent to averaging over specific fractions of year only, 
it is not possible to test such approximations without numerical experiments,
the most relevant case being the interplay between the change 
in flux due to the time dependence in the Sun - Earth distance 
$\RSE(t)$\ and the time dependence in the Earth orbital velocity
\cite{ArticleI}. 
In conclusion,
all the results presented here are obtained using the EOA, which assures
an accuracy for \APeTw\ largely better than $1\%$\ for each kind of sample
and in what regards the solar motion. 

\section{Results}
At the basis of each time averaged probability \APeTw\ is the instantaneous
one, \PeTw.
Figure \FigPeTw\ shows an example of the probability \PeTw\ for $\SdTvS = 0.01$
as a function of \hath\ and of the {\em Resonance Density} in the Earth \rhoR\
(gr/cm$^3$)
\cite{ArticleI}: $\rhoR \Ye = 6.57 \times 10^6 (\dms(\eV^2) /\Enu(\MeV)) \CdTv$.
There are two main peaks associated to resonance transitions in the
core, {\tt C}, and in the mantle {\tt M} \cite{ArticleI}. 
In addition, the spherical symmetry of the Earth implies that solar neutrinos 
will cross the resonance region twice. 
Since the neutrino oscillation length in matter,
for solar neutrinos which undergo resonance transitions in the Earth, is 
of the order of the Earth radius, interference terms 
in \PeTw\ lead to oscillatory dependence of \PeTw\ on \hath\ for a fixed 
\rhoR\ \cite{ArticleI,Baltz:Weneser:1994}. 

Increasing \SdTvS, for 
$\SdTvS < 0.1$, the qualitative properties of this dependences do not 
change significantly. The two peaks scales accordingly to \SdTvS, with 
{\tt C} increasing faster than {\tt M} reaching a maximum
value and begins to decrease while {\tt M} is still increasing.
As a consequence, at values of $\SdTvS \, \lsim \, 0.13$\ 
most of the \daynight\
asymmetry is generated by the MSW effect in the core.
At larger \SdTvS\ values the behaviour is more complicate, new peaks appears
and (for $\SdTvS \ge 0.3$) the bulk of the Earth transition is in the mantle.

As Fig. {\FigAPeTw}a illustrates, the separation in \core\ and
\mantle\ samples is very effective in enhancing \APeTw.
The enhancement is particularly large for small mixing angles for which
\APeTwC\ can be up to six times bigger than \APeTwM\
(for $\SdTvS \ge 0.3$\ the core - enhancemente is only $30\%\div40\%$).
This suggests the possibility of a corresponding enhancement of the
$e^{-}$-spectrum distortions
as well as of the energy integrated event rate. 
To have an enhancement in the averaged probability \APeTw\
is not enough to ensure an enhancement in the \daynight\ asymmetry,
especially at small mixing angles.
The upper ``window'' of Fig. {\FigAPeTw}b is a combined plot of \Ps\ 
(dotted line), \PeTw\
(dashed line) and \PTot\ (solid line) for \core. 
In the lower ``windows'' there is a plot of the ratio:
\getup\bec\beq
          \AsymPs(\EDms) \equiv 2~\frac{\PTot(\EDms)  - \Ps(\EDms) }
{\PTot(\EDms)  + \Ps(\EDms) }
\eeq\eec

\noindent
which is the \daynight\ asymmetry for the total survival probability.
The large peak on the left side of the figure  is an artifact 
of the presence of the adiabatic
minimum in the probability \Ps.
The bulk of the \daynight\ asymmetry comes
from the second peak in the figure.

Figure \FigAPeTw\ is an example of how each probability
contributes in \PTot\ and in the related asymmetry.
One of the important features that has to be taken into account in the
discussion of \PTot\ is the position of the peaks in \APeTw\ with respect
to the value of \EDms\ at which $\Ps = 0.5$\ because it controls the sign
and the magnitude of the \daynight\ effect. It follows from eq.
(\ref{eq:neutrino:survival:proba}) that the asymmetry is zero each time
$\Ps = 0.5$\ irrespective of the value of \APeTw.  
If $\APeTw > \STvS$, and 
$\Ps < 0.5$ ($\Ps > 0.5$), 
the asymmetry is positive 
(negative) which means the Earth effect {\em increases}
({\em reduces}) the night event rate \cite{ArticleI}.
The presence of a zero Earth effect 
can reduce the magnitude of the \daynight\ asymmetry.
The negative \daynight\ effect is relevant for $\SdTvS < 0.004 \div 0.006$.

The definitions for the
recoil-\electron\ spectrum
with (without) MSW effect
$\Ses(\Te)$\ 
(\SeZr(\Te)), 
\Te\ being the recoil-\electron kinetic energy
and $s=D$, $N$, $C$, $M$;
and for the energy integrated event rates with (without) MSW effect 
$\Res(\TeTh)$\ 
$(\ReZr(\TeTh))$, \TeTh\ being the recoil-\electron 
threshold kinetic energy, in terms of the survival probability \PTot,
are reported in \cite{ArticleII}
Here it is enough to recall the definition of the distortion for 
the recoil-\electron spectrum due to the MSW effect:
$\deltaSs(\Te) = {\Ses(\Te)}/{\SeZr(\Te)}$, 
$s = \mbox{$D$, $N$, $C$, $M$}$,
%
and the definitions for the \daynight\ asymmetries for the recoil-\electron 
spectrum $\AsymSs(\Te)$\ and
the energy integrated event rate $\AsymRs(\TeTh)$:
\getup\bec\beq
     \AsymSs(\Te) = 2 \,\, \frac{\Ses(\Te)-\SeD(\Te)}{\Ses(\Te)+\SeD(\Te)},
      \mbox{\hspace{1cm}}
     \AsymRs (\TeTh) = 2 \,\, \frac{\Res - \ReD}{\Res+\ReD},
\eeq\eec

\noindent
with $s=D$, $N$, $C$, $M$.
Both these observable quantities  are solar model independent,
and are related to $\AsymPs(\Enu/\dms)$,
$s=D$, $N$, $C$, $M$, $DC$.
For fixed values of \SdTvS\ and \TeTh:
\getup\bec\beq\label{eq:relaz:asym}
\left.
\begin{array}{cll}
    \maxim(\AsymRs) &\approx &
    \frac{\maxim(\AsymPs(E/\dms) \Ps_{,max}}
         {\alpha - \beta  \,\, \maxim(\AsymPs(E/\dms) + \Ps_{,max}},\\
&&\\
    \AsymSs(E-0.5\MeV) &\approx &
    \frac{\AsymPs(E/\dms) \Ps(E/\dms)}
         {\alpha - \beta \,\, \AsymPs(E/\dms) + \Ps(E/\dms)},\\
\end{array}
\right. 
\end{equation}\end{center}

\noindent
where $s =D,\, N,\, C,\, M,\, DC$, $\alpha=0.190476$, $\beta=0.0952381$.
In the upper formula $\AsymRs$\ is regarded as a function of \dms,
$\Ps_{,max}$\ is the value of $\Ps(\Enu/\dms)$\ computed for the $\Enu/\dms$\
value which corresponds to the $\AsymPs(\Enu/\dms)$\ maximum
$\maxim(\AsymPs)$.
In the lower formula \dms\ is fixed, while it is assumed:
$E \cong \Enu \cong \Te+0.5\MeV$. 
Equations (\ref{eq:relaz:asym}) 
overestimate the numerical results by a factor between 
$3\%$\ and $40\%$. The error decreases increasing \SdTvS.

One example of the predicted recoil - e$^{-}$ spectrum distortion in the Sun, 
and one example of \daynight\ asymmetries in the spectrum are shown in 
Figs. \FigSDist\ (see \cite{ArticleII,SKDNII:spectrum}\ for a larger set of 
examples) both the figures being computed for $\Ye=0.467$.
The enhancement of the spectrum distortions and of the \daynight\
asymmetry in the \core\ sample is clearly seen,
especially for $\SdTvS \lsim 0.013$.
The differences between the spectra computed for $Y_e=0.467$\ and for 
$Y_e = 0.5$, are largely negligible for 
the expected experimental accuracy.


  The differences between the \DAY\ and the \night\ (\mantle) sample
\electron\ - spectra are hardly observable, but for $\SdTvS \geq 0.008$\
and many values of \dms\ from the ``conservative'' MSW solution region,
$\AsymSC(\Te)$\
may reach a value of $10\%$\ 
and even larger ($20\% - 30\%$). So, its measurement can further 
constrain \dms\ and \SdTvS\ values.

Figures \FigContour, \FigContourFive\
represent the iso-(\daynight) asymmetry contour plots for the 
\night, \core\ and \mantle\ samples. 
Shown are also the ``conservative'' regions of the two MSW solutions
of the solar neutrino problem (dashed lines) as well as 
the solution regions obtained in the reference solar model
\cite{BP95}\ (inner thick dashed lines). 
Also the iso-(\daynight) asymmetry contours for the 
\night\ and \core\ samples shown in 
Figs. \FigContour\ and \FigContourFive\
have been obtained both for $Y_e = 0.467$ (solid lines) and for
$Y_e = 0.5$ (dash-dotted lines).
For given \dms\ and \SdTvS\ the differences 
between \AsymRN\ or \AsymRM\ 
and \AsymRC\ are larger in the NA region.
 The magnitude of the \daynight\ asymmetry 
in the AD region does not change 
significantly with the change of the sample.
   In contrast, in the NA region the \daynight\ asymmetry 
in the \core\ sample
is enhanced by a factor of up to six. The reduction in 
statistics for this sample  
reduces it to an effective 2.3 enhancement.
From Fig. \FigContourFive\ it comes 
that the \core\ selection is a very effective method to enhance the 
\daynight\ asymmetry despite the loss of statistics. 

The difference between the values of the \core\ sample 
asymmetry \AsymRC\ calculated for $Y_e = 0.467$ and for
$Y_e = 0.5$ depends strongly on the values of \dms\ and \SdTvS. 
However, Figs. \FigContourFive\ show that 
the main results illustrated here are stable against the 
Earth model uncertainty
\cite{ArticleII}.
  
   Because of the  large differences between the core and the 
mantle structures, the 
\mantle\ and \core\ subsamples provide two independent measurements of 
the \daynight\ effect. These can be combined to constrain better 
the neutrino parameters \dms\ and \SdTvS, 
utilizing the full statistics of the \night\ sample, especially in NA region
\cite{ArticleII}. 

Figs. \FigContourFive\ suggests also the possibility to use the \daynight\ 
effect to limit the \BHt\ neutrinos total flux \cite{ArticleII}.

\section{Conclusions}
In this talk the possibility of a core-enhancement for the \daynight\ effect 
in the \SK\ detector, proposed by example in \cite{Baltz:Weneser:1994}, 
was discussed in detail.
It was remarked the need of a precise apparent solar motion 
reconstruction to produce the accurate \daynight\ effect predictions
required to study this subject.
A computer code which fulfills this requirement was presented.
Predictions for \SK\ detector observable quantities  were obtained.
It was shown that:
I). The \core\  sample \daynight\ asymmetry can be enhanced up to factor six
   compared to the \night\ and \mantle\ \daynight\ asymmetries. 
   The enhancement reduces to a factor 2.3 when the loss in statistics
   is properly accounted for. 
   If \SK\ will be able to reach an accuracy of some percent in measuring
   the \core\ \daynight\ asymmetry it is reasonable
   to expect it will be able to test this region through the \daynight\ 
   effect. 
II). The enhancement may be relevant for the recoil-\electron\ spectra
    distortion and the recoil-\electron\ spectra \daynight\ asymmetry. 
III). The detection (or lack of detection) of a small \daynight\ asymmetry 
    may constraint the total \BHt\ neutrino flux from the Sun
    in a solar model independent way (presently this flux is uncertain for
    a factor two).
IV). These results are qualitatively robust against the Earth model 
    uncertainties.

\section*{Acknowledgments}
The author is deeply indebted with Prof. W. Hampel and the organizing 
committee of the {\em Fourth International Solar Neutrino Conference} for the 
kind concession of a financial support to attend to the conference.
This research was partially supported by S.I.S.S.A. Trieste,
Italy, research contracts: 4060 and 5566.


\end{document}